\documentclass[preprint2]{aastex}
 
\usepackage{color}

\usepackage{lineno}
\bibpunct{(}{)}{;}{a}{}{,}

\shorttitle{FAST polarization mapping of the SNR G166.0$+$4.3}

\shortauthors{Xiao et al.}

\begin{document}

\title{FAST polarization mapping of the SNR VRO~42.05.01 }

%\volnopage{Vol.0 (200x) No.0, 000--000}      %%preserved for Editor. DOn't remove!
\setcounter{page}{1}

%\offprints{L. Xiao}

\author{
L\sc{i} X\sc{iao},\altaffilmark{1,2,3}
M\sc{ing} Z\sc{hu},\altaffilmark{1,2,3}
X\sc{iao}-H\sc{ui} S\sc{un},\altaffilmark{4}
P\sc{eng} J\sc{iang} \altaffilmark{1,2,3}
C\sc{hun} S\sc{un} \altaffilmark{1,2,3}       
}

\altaffiltext{1}{National Astronomical Observatories, CAS, Beijing 100012, China; Email: xl@nao.cas.cn}
\altaffiltext{2}{Key Laboratory of FAST, NAOC, Chinese Academy of Science, Beijing 100012, China}
\altaffiltext{3}{Guizhou Radio Astronomical Observatory, Guizhou University, Guiyang 550000, China}
\altaffiltext{4}{Department of Astronomy, Yunnan University, and Key Laboratory
of Astroparticle Physics of Yunnan Province, Kunming, 650091, China \\}

%  \email{xl@nao.cas.cn}

%\date{Received / Accepted}

\begin{abstract}
We have obtained the polarization data cube of the VRO~42.05.01 supernova remnant at 1240~MHz using the Five-hundred-meter Aperture Spherical radio Telescope (FAST).
Three-dimensional Faraday Synthesis is applied to the FAST data to derive the Faraday depth spectrum. The peak Faraday depth map shows a large area of enhanced foreground RM 
of $\sim$60~rad~m$^{-2}$ extending along the remnant's ``wing'' section, which coincides with a large-scale HI shell at $-20$~km~s$^{-1}$.
The two depolarization patches within the ``wing'' region with RM of 97~rad~m$^{-2}$ and 55~rad~m$^{-2}$
coincide with two HI structures in the HI shell. 
Faraday screen model fitting on the Canadian Galactic Plane Survey (CGPS) 1420~MHz full-scale polarization data reveals
a distance of $0.7-0.8d_{SNR}$ in front of the SNR with enhanced regular magnetic field there.
The highly piled-up magnetic field indicates that the HI shell at $-20$~km~s$^{-1}$ could originate from an old evolved SNR. 
%and probably has affected the regular magnetic field in the surrounding medium of VRO~42.05.01.

\end{abstract}


\maketitle

\section{Introduction}
A supernova remnant shock compresses the magnetic field in the interstellar medium (ISM) with a factor of four or more in the evolved stage~\citep{fr04,hvb12},
which generates enhanced synchrotron emission and retains the information of ambient interacted medium.
Radio polarization observations directly detect the perpendicular magnetic field configuration across supernova remnants (SNRs),
and give the line-of-sight component of the magnetic field by measuring the Faraday rotation of the polarization angles at different frequencies.
It provides important magnetic field configuration information for understanding the dynamics of shock, post-shock gas, and the interaction with the ISM.

VRO~42.05.01 (G166.0$+$4.3) is a shell-type SNR with a peculiar morphology consisting of a semicircular ``shell'' and a
triangular ``wing'' structure~\citep{ppl85,plr87}. It has been observed in various bands to investigate the evolution in the 
inhomogeneous medium with two different densities (Summary review in \citet{adz19}).
Radio observations from low to high frequencies have indicated different indices and compression ratios in these two sections~\citep{lt05,ghr11,avi19,xzs22}.
The polarized emission has been observed in the Canadian Galactic Plane Survey (CGPS) at 1420~MHz, with most regions depolarized~\citep{kff06}.
A large-scale HI shell at $-20$~km~s$^{-1}$ with the corresponding footprint on the $U/Q$ map extends across the remnant, probably acting as a
foreground Faraday screen~\citep{kl04}. The remnant has also been observed in the Sino-German Galactic plane survey at 4.8~GHz~\citep{ghr11}.
The magnetic field vectors in the two ``wing'' shock shells show different deviations from tangential to the shock front, indicating different rotation measures.   

In this paper, we present the polarization observation of VRO~42.05.01 at 1240~MHz with the Five-hundred-meter Aperture Spherical radio Telescope (FAST). 
FAST has the advantage of a large aperture with high resolution and high sensitivity~\citep{jth20} and has produced good scientific outputs~\citep{qys20}.
The 19-beam wide band multi-channel polarization receiver from 1050$-$1450~MHz enables it to measure the polarization of extended SNRs
at multi frequencies simultaneously~\citep{sgr22,grs22}. Through the Faraday rotation synthesis technology~\citep{bd05}, 
we can efficiently obtain the Faraday depth spectrum and retrieve the polarization properties towards VRO~42.05.01.

The paper is organized as follows. The observation and data reduction are described in Sect. 2. 
The comparison of the FAST polarization map with the CGPS map is shown in Sect. 3.1. 
The Faraday rotation synthesis and depolarization analysis are presented in Sect. 3.2 and 3.3. In Sect. 4, we make a discussion on the 
depolarization feature in the southern wing region. Conclusions are summarized in Sect. 5.

\section{Observations and Data reduction}
%\subsection{Observations and data processing}

The FAST polarized continuum mapping of VRO~42.05.01 was observed twice using the multibeam on-the-fly mode (2RA + 2DEC) %along the R.A. and decl. direction
in 2020 April and 2021 February, respectively. The observation setting and parameters are listed in paper I~\citep{xzs22}. 
3C138 is observed as the calibrator. The spectral backend with a 65536 channel covering the 1050$-$1450~MHz band recorded 
the four polarization outputs ($I_{1}$, $I_{2}$, $U$, $V$) from the FAST 19-beam linear feeds.
After intensity calibration, the data were converted to antenna temperature (K~$T_{a}$),
and the Stokes parameters are derived as $I = I_{1}+I_{2}$, $Q=I_{1}-I_{2}$, $U$ and $V$.
The leakage caused by the mismatch between the amplitudes and phases of the gains of the two linear feeds is small and
corrected using the 10~K injected reference signal injected every two seconds following the method in~\citet{smg21}. 

For the $Q$, $U$, and $V$ data along the R.A. or decl. direction, a baseline correction was first made by linearly fitting the two ends of each scan and 
subsequently subtracted to remove unrelated large-scale emissions. Radio frequency interference (RFI) caused by communication satellites and navigation satellites 
(1160-1280~MHz) and other interference were flagged the same as that in $I$ in the frequency zone.  
Then we converted the antenna temperature to the main-beam brightness temperature (K~$T_{B}$) by dividing the main beam efficiency ($\eta_{b}$) 
calculated from the calibrator~\citep{smg21}, and smoothed each channel map to a common 4$\arcmin$ angular resolution. 

Finally, all four $Q$ maps and $U$ maps were binned every 20 channels to improve the noise level and weaved together to destripe the scanning effects 
in the Fourier domain~\citep{eg88}. We averaged the data over the full FAST band by taking medians of all the frequency channels to form the combined $Q$ and $U$ maps. 
The polarized intensity $PI$ corrected from the positive noise bias was calculated as $PI = \sqrt{Q^{2}+U^{2}-(1.2\sigma)^{2}}$~\citep{wk74},
where the rms noise $\sigma$ measured from the average combined $Q$, $U$ maps are both about 4~mK T$_{b}$. 
The polarization angle $\psi$ is derived as $\psi = \frac{1}{2}\arctan2(U, Q)$. 
%$\psi = \frac{1}{2}\arctan\frac{U}{Q}$.  

\begin{figure*}[!hbt]
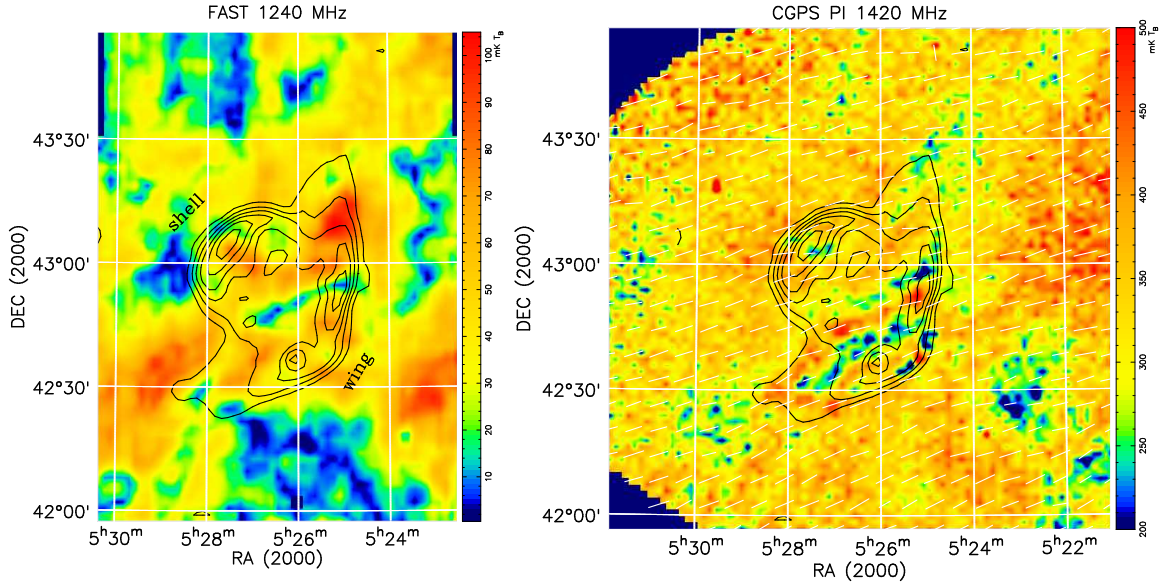

\begin{center}
\includegraphics[angle=-90,width=0.4\textwidth]{G166.pi.noB.ps}
\includegraphics[angle=-90,width=0.52\textwidth]{cgps.pipa.ps}
\caption{ Left panel: The FAST full bandwidth polarization intensity map of VRO~42.05.01 at a central frequency of 1240~MHz.  
The angular resolution is 4$\arcmin$. Contours show the total intensities of VRO~42.05.01 start at 350~mK
and increase by a step of 300~mK. 
Right panel: The CGPS 1420~MHz polarization intensity map of VRO~42.05.01 with the large-scale polarized emission restored from Effelsberg and 
the DRAO 26-m single antenna telescopes. The angular resolution is 1$\arcmin$. The contours are the same as the left panel. 
The bars show the orientation of the magnetic field B, with lengths proportional to the CGPS polarization intensity.
}
\label{pi}
\end{center}
\end{figure*}

\section{Results}
\subsection{Comparison with the CGPS 1420~MHz polarization map}
The FAST full-bandwidth 1240~MHz polarization intensity map of VRO~42.05.01 derived from the averaged $Q$ and $U$ data is presented in Fig.~\ref{pi}. 
There is a weak polarization emission counterpart for the shock ``shell''.
The polarized patches in the center of VRO~42.05.01 with a comparable 10$-12$~mK T$_{b}$ intensity with that in the outer region,
might originate from the foreground diffuse polarization emission. A zero-polarized intensity canal structure with a length of 27$\arcmin$
passes across the wing region. It has a width of about $4\arcmin$, similar to the beam size, and might be caused by different polarization angles 
of the polarized patches on each side.

For comparison, the combined Canadian Galactic Plane Survey (CGPS) 1420~MHz polarization intensity map of VRO~42.05.01~\citep{lrr10} is shown in
the right panel of Fig.~\ref{pi}. It has an angular resolution of $\sim 1\arcmin$, and has restored the large-scale polarized emission 
by combing data from Effelsberg and the Dominion Radio Astrophysical Observatory (DRAO) 26-m single antenna telescopes. 
Polarized emission is present only in the shell peak and the tip edge of the triangle wing region in the CGPS synthesis map (Fig.13 in ~\citet{kff06}),
which turns into voids after baselevel restoration.
The depolarization feature in the southern wing region (two depolarization patches separated with a filament) is also identical in a way of opposite values
to the FAST 1240~MHz polarization intensity map.
The canal structure becomes polarized filaments in the CGPS 1420~MHz map, while the polarization patches on both sides turned into depolarized regions.

The large-scale emission component is important for the interpretation of polarized structures caused by Faraday rotation
in the interstellar medium~\citep{r06}. The FAST 1240~MHz polarization data lacks the large-scale polarized emission due to the baselevel subtraction. 
After adding the large-scale polarization vectors in the CGPS map, the polarized canals become polarized filament emission. 
The polarized patches in the southern wing region become depolarized regions, 
indicating there exists Faraday rotation effect. The polarization angle of the background polarization emission has been modulated 
and canceled the foreground polarization emission to cause a depression in the polarization intensity (detail discussion in Sect. 4.1).

\begin{figure*}[!hbt]
\begin{center}
\includegraphics[angle=0,width=0.9\textwidth]{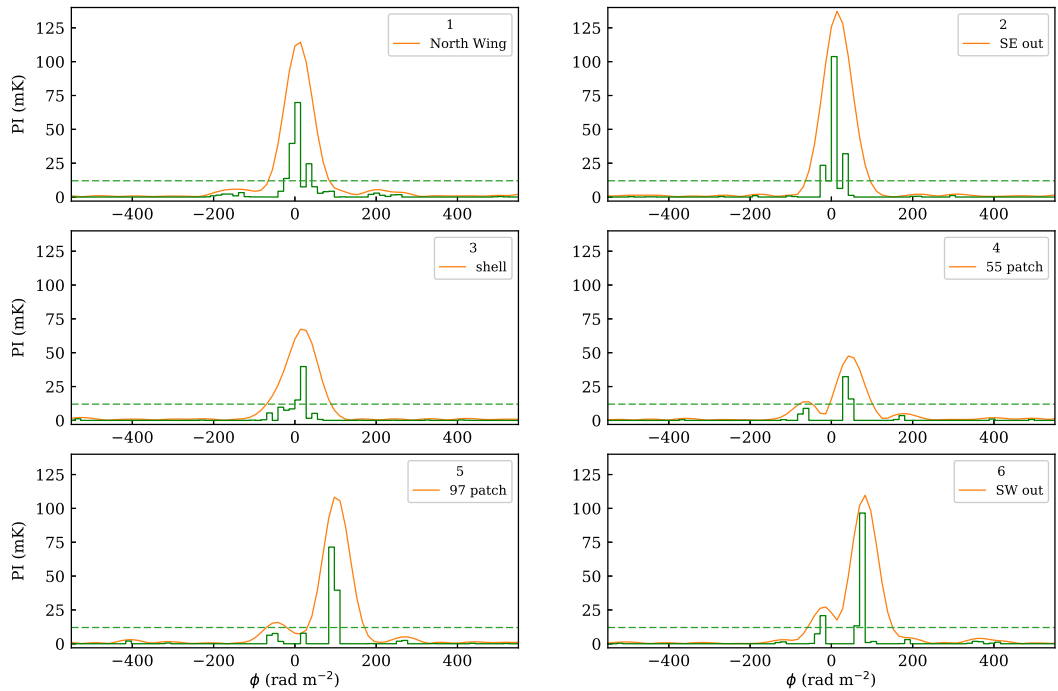}
\caption{The Faraday depth spectra and fitted Faraday depth $\phi$ components (green) towards polarization patches in the region of VRO~42.05.01 as
marked in the maxPI map of Fig.~\ref{maxpi}. 
}
\label{FDF}
\end{center}
\end{figure*}

\subsection{the Faraday Synthesis analysis}

We used the widely used RM synthesis to reconstruct the Faraday depth spectrum from the frequency cubes of $Q$ and $U$~\citep{bd05}.
The method is based on that the observed polarization intensity $P(\lambda^{2})$ and the Faraday dispersion function $F(\phi)$ are 
Fourier transform pairs~\citep{b66} %(equation.~\ref{RMsyn})

\begin{equation}\label{RMsyn}
P(\lambda^{2})= Q(\lambda^{2})+iU(\lambda^{2})=\int F(\phi)e^{2i\phi\lambda^{2}}d\phi,  
\end{equation}
where $\lambda$ is the wavelength and the $\phi$ is the Faraday depth defined 
as $\phi=0.812\int n_{e}B_{||}dr$, integral of the thermal electron density with the magnetic field along the line-of-sight from the source to the observer.
By fitting the peak of the Faraday depth spectrum, we can obtain the Faraday depth integrated towards the source, 
which is equivalent to the rotation measure (RM).

We applied the RM synthesis to the FAST data of VRO~42.05.01 for each pixel using the RM-Tools package~\citep{pvw20}. 
The algorithm returns the peak Faraday depth $\phi_{peak}$, the peak polarized intensity $F(\phi_{peak})$, and the polarized intensity cube along Faraday depth $F(\phi)$.
For the FAST frequency range and sampling, the width of the RM spread function is about 90~rad~m$^{-2}$, and the Faraday depth has a
step interval of 14~rad~m$^{-2}$. 
We displayed Faraday spectra and fitted peak components towards six polarized patches as marked in the maxPI map in Fig.~\ref{FDF}.
Most of the Faraday spectra show one peak above 3$\sigma_{PI}$ level. The spectra towards the polarized patches in the northern wing and
in the southeast outside peak at 0~rad~m$^{-2}$ indicate, that they come from the foreground and the polarized emission of the remnant is fully depolarized. 
The spectrum in the shell shows a peak of 14~rad~m$^{-2}$. The spectra towards the polarized patches in the southern wing (4, 5) and in the southwest outside (6) 
show large RM peaks of 55, 97, and 97~rad~m$^{-2}$, respectively. The spectrum 6 also shows another weak negative RM component, which might
come from a foreground local feature. Here we just discuss the positive RM.

\begin{figure*}[!hbt]
\begin{center}
\includegraphics[angle=0,width=0.99\textwidth]{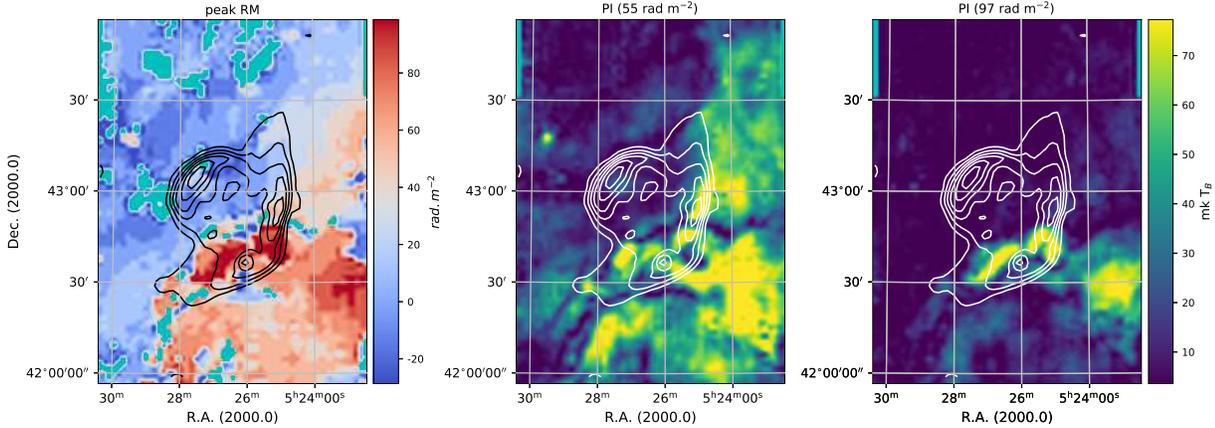}
\caption{Left: The peak RM map reconstructed from the Faraday synthesis method for VRO~42.05.01. Pixels with maxPI values below 30 mK have been set to nan.
Middle and right: the polarized intensity at 55~rad~m$^{-2}$ and 97~rad~m$^{-2}$ to display the large RM distribution separately.  
}
\label{rm_syn}
\end{center}
\end{figure*}
 
\begin{figure}[!hbt]
\begin{center}
\includegraphics[angle=-90,width=0.4\textwidth]{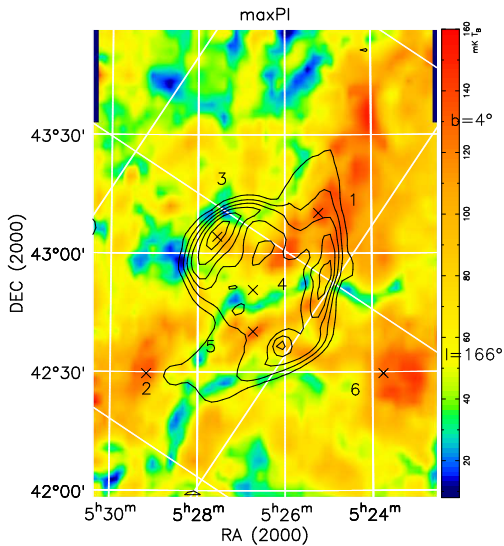}
\caption{The FAST maximum polarized intensity of VRO~42.05.01 at 1240~MHz derived from the Faraday synthesis method.
}
\label{maxpi}
\end{center}
\end{figure}

The peak Faraday depth $\phi_{peak}$ map towards VRO~42.05.01 is presented in Fig.~\ref{rm_syn}.
It shows the large Faraday depth  area ($\phi>$ 60~rad~m$^{-2}$) in the southwest extending along the wing region (also the Galactic longitude direction).  
The depolarization patches within the wing region in the CGPS map of Fig.~\ref{pi} correspond to a maximum Faraday depth of 97~rad~m$^{-2}$ and 55~rad~m$^{-2}$, respectively.
The polarized intensity at these two values in the Faraday depth cube $F(\phi)$ are shown separately in Fig.~\ref{rm_syn}.

We presented the maximum polarized intensity map in Fig.~\ref{maxpi}. The canal and polarized patches in the wing region
are almost the same as in the polarization map averaged from $Q$ and $U$. However, in the southern and western edges, the maxPI map shows some weak polarized patches,
which are depolarized in the averaged polarization map. Considering the large foreground RM of 60~rad~m$^{-2}$, 
averaging over all the frequency channels might have caused some bandwidth depolarization towards these regions.

\subsection{Depolarization in the shell}
In most regions of VRO~42.05.01 the polarization has been smeared out. Only weak polarized emission is detected in the shell region both 
in FAST 1240 and CGPS 1420~MHz maps. 
For a medium where synchrotron emission and Faraday rotation co-exist, depth depolarization is the main mechanism to cause depolarization.
Polarized emission at different depths experiences 
different Faraday rotations, which reduce or even cancel the polarization intensity when added together.
Following~\citet{sbs98}, it is wavelength-dependent and the depolarization $P_{\lambda}$, the ratio of the observed polarization percentage $PC$ at 
$\lambda$ to intrinsic $PC$ is related to RM as 
\begin{equation}\label{energy}
P_{\lambda}={\rm sin}(2|RM|\lambda^{2})/ 2|RM|\lambda ^{2}
\end{equation}

\begin{figure}[!hbt]
\begin{center}
\includegraphics[angle=-90,width=0.46\textwidth]{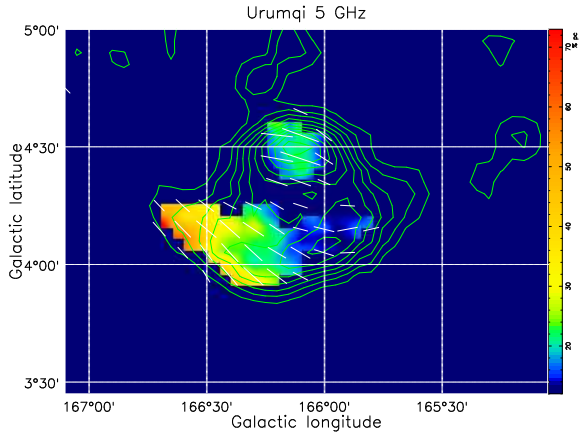}
\caption{ The Urumqi 4.8~GHz polarization percentage map of VRO~42.05.01.  
The image has an angular resolution of 9.5$\arcmin$. Contours show the total intensities at 4.8~GHz start at 6~mK, 
and increase by a step of 5~mK. The bars show the orientation of the magnetic field B, with lengths proportional to the polarized intensity. 
}
\label{6cmpc}
\end{center}
\end{figure}

We presented the Urumqi 4.8~GHz polarization percentage map~\citep{ghr11} in Fig.~\ref{6cmpc}. At 4.8~GHz, the depth depolarization is 
negligible and it traces approximately the intrinsic polarization emission. The polarization percentage $PC$ in the shell region is about 20\%.
The wing region has a different $PC$ distribution, with the southern section having a higher $PC$ of 24\%, and a lower $PC$ of 15\% in the northern area.
By comparing the polarization degree in the shell at 1240~MHz ($PC\sim 3\%$) with that at 4.8~GHz, we obtained a relative depolarization $PC_{1.24}/PC_{5}$
of 0.85. It corresponds to an intrinsic RM of 20~rad~m$^{-2}$ in the shell peak. The positive sign of RMs is adopted
based on the deviation of polarization angles at 4.8~GHz. The value is comparable with the Faraday synthesis result in Fig.~\ref{rm_syn}.

\begin{figure}[!hbt]
\begin{center}
\includegraphics[angle=-90,width=0.46\textwidth]{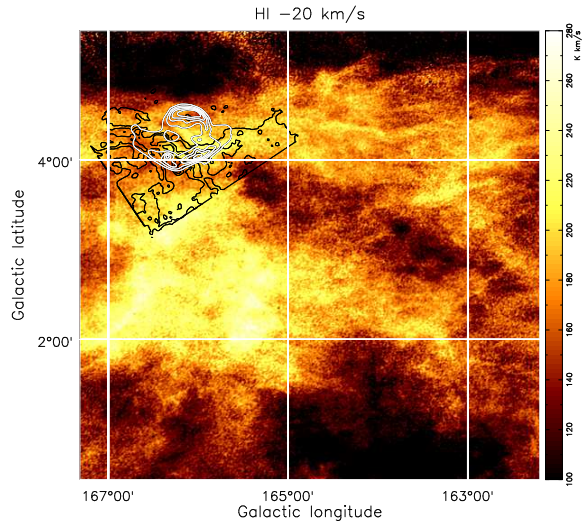}
\caption{The CGPS integrated HI map with 3 channels at $-20$~km~s$^{-1}$ towards VRO~42.05.01. White contours show the CGPS total intensity of VRO~42.05.01 smoothed to 4$\arcmin$, 
ranging from 4.8~K in a step of 240~mK. Black contours show the Rotation measure map at 55~rad~m$^{-2}$ at 26 and 60~mK level.
}
\label{hi_rm}
\end{center}
\end{figure}

\section{Discussion}
\subsection{depolarization feature caused by Faraday screens}
VRO~42.05.01 was reported to overlap with a partial large-scale HI shell with a radius of 4$\degr$ at $-20$~km~s$^{-1}$ in the CGPS polarization survey~\citep{kl04}.
As shown in Fig.~\ref{hi_rm}, the HI shell seems to go across and have separated the ``shell'' and ``wing'' region.
This is consistent with the RM peak distribution derived from the FAST polarization data of Fig.~\ref{rm_syn}.
The large foreground RM component at 55~rad~m$^{-2}$ in the wing section also extends along the HI shell.

\begin{figure}[!hbt]
\begin{center}
\includegraphics[angle=-90,width=0.46\textwidth]{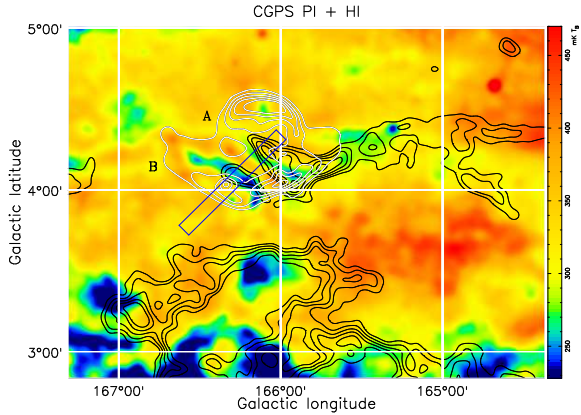}
\caption{The CGPS 1420~MHz restored polarization map of VRO~42.05.01 smoothed to 4$\arcmin$. Black contours show the CGPS $-20$~km~s$^{-1}$ HI channel map 
smoothed to 4$\arcmin$, ranging from 76.3~K in a step of 2.5~K. White contours is the same as Fig.~\ref{hi_rm}.
}
\label{pi_hi}
\end{center}
\end{figure}

The depolarized patches in the wing region of VRO~42.05.01 seem to coincide with two HI sheet structures in the HI shell.
As revealed in the CGPS restored polarization intensity map smoothed to 4$\arcmin$ in Fig.~\ref{pi_hi}, 
the lower end of the depolarized patch (B) with a width of 5.0$\arcmin$ overlaps with the HI cloud below and extends upward.
The depolarized arc (A) with a width of 2.4$\arcmin$ goes along the left periphery of the upper HI cloud and gradually extends to coincide with the right side of the cloud.
It seems that the HI sheet structures have enhanced rotation measures and acted as Faraday screens, which rotated the background polarized emission and
caused depolarization after adding up with the foreground emission.

\begin{figure}[!hbt]
\begin{center}
\includegraphics[angle=0,width=0.48\textwidth]{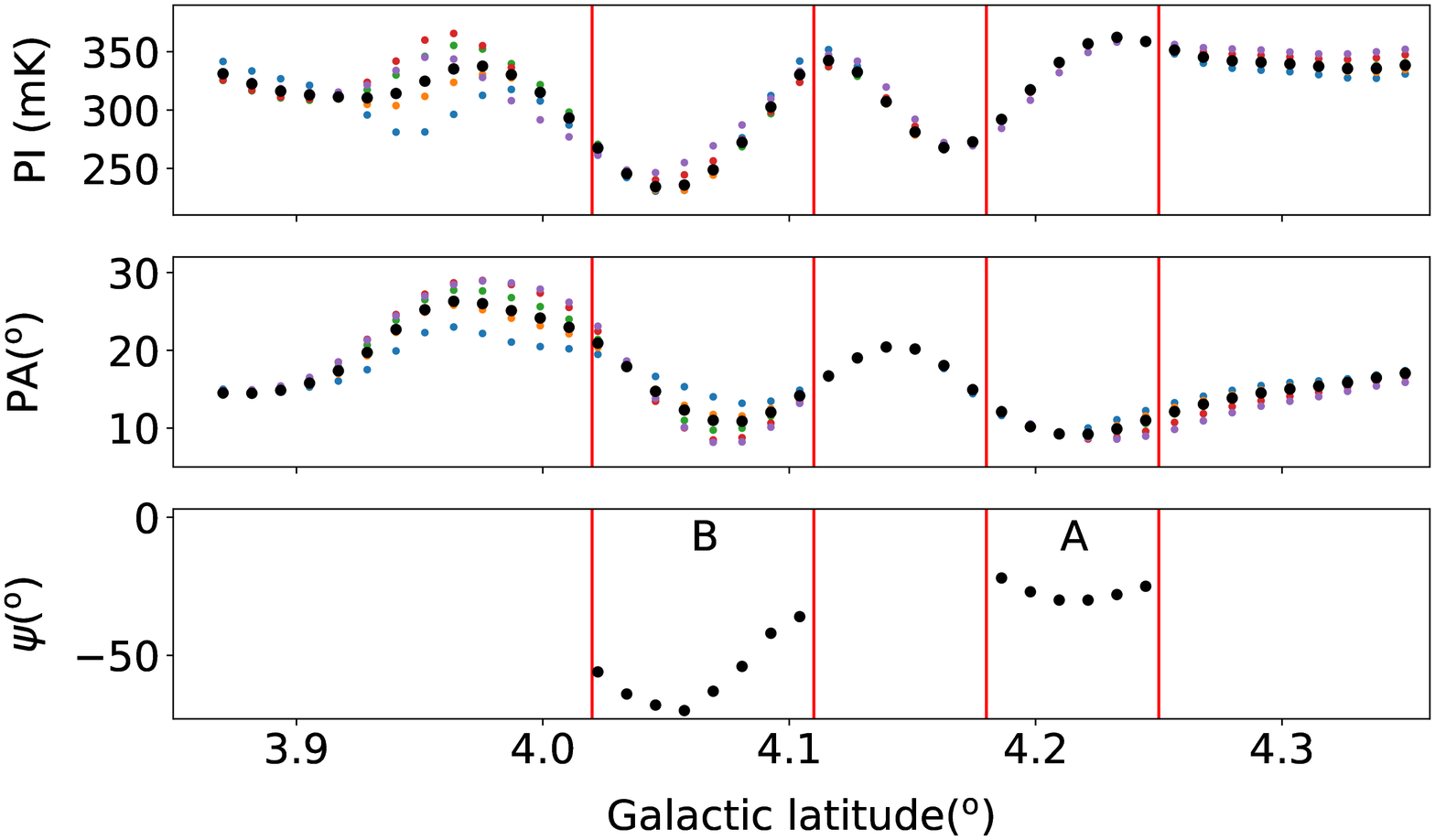}
\caption{The distribution of the CGPS $PI$, $PA$ along the slice in the southern wing region of VRO~42.05.01.
The fitted rotation angle $\psi$ according to the Faraday screen model are $-$30$\degr$ and $-$70$\degr$ with an ambiguity of n$\pi$ (n=0, 1, 2, $\cdots$). 
}
\label{fs}
\end{center}
\end{figure}

We tried to apply the Faraday screen model to the CGPS 1420~MHz polarization map. We extracted data along the slice in Fig.\ref{pi_hi}.
The $PI$ and $PA$ distribution in the rectangle region along the slice direction is presented in Fig.~\ref{fs}. 
The distance of VRO~42.05.01 is 3.24~kpc from the optical extinction of background stars~\citep{zjl20}.
The total depolarization in the wing region indicates that all detected polarized emission at 1.24~GHz comes from the foreground,
closer to us than the remnant. It is consistent with the CGPS polarization horizon estimation of 3.0~kpc at $l<140\degr$~\citep{kl04} from 
depolarization properties of HII regions and SNRs. Assuming that the Galactic disk has a uniform polarized emissivity and rotation measure distribution, 
we fitted the ratio of polarization intensity and polarization angle ($PI_{on}/PI_{off}$, $PA_{on}/PA_{off}$) through the Faraday screens with 
that in the off surrounding region according to the formulas in~\citep{srh11}, to obtain their relative distances with the remnant and the rotated angles. 
The fitted distances of these two Faraday screens are 0.7$-$0.8$d_{SNR}$.
The depths of the Faraday screens assuming equal to the mean width are then estimated about 1.8 and 3.8~pc.
The fitted maximum rotated angles shown in Fig.~\ref{fs} are about $-$30$\degr$ and $-$70$\degr$, with an ambiguity of $\pm n\pi$ (n=0, 1, 2, $\cdots$).
Based on the RM synthesis results of 55 and 97~rad~m$^{-2}$ towards these two regions, the corresponding rotated angles are 150$\degr$ and 290$\degr$,   
respectively.

The non-detection of radiation from the Faraday screens in the total intensity gives a limit on the electron density.
According to ~\citet{rw04}, the contribution of warm thermal gas to the FAST 1240~MHz brightness temperature could be estimated as $T=T_{e}\tau=2.26n_{e}^{2}L$~mK,
where $\tau$ is the opacity, $L$ is the radiation depth in pc and $T_{e}$ is the electron temperature taken to be 8000~K.
Take $5\times rms$~(12~mK) in the total intensity as the non-detection level, the upper limits of electron density in the Faraday screens are derived as
1.7 and 1.2~cm$^{-3}$.
The resulting lower limits for the regular magnetic field strength along the line of sight are about 22 and 26~$\mu$G.
However, the value is inversely proportional to the square root of the depth $L$ and thus largely depends on the geometry of the depolarized screens. 
If the depolarized structures are sheets or a shell periphery with ten times large depths, the magnetic field upper limit will go down to $6-8$~$\mu$G.

The large-scale HI shell at $-20$~km~s$^{-1}$ was suggested to be swept up by an old SNR evolved into the dissipation phase~\citep{kl04}. 
The thermal and radio emission in the shocked shell has declined and the interior temperature drops, while the piled-up regular magnetic field and HI shell remain. 
A similar case of HI bubbles associated with old remnants has been reported in~\citet{xz14}.

\subsection{The RM in the wing region}
The large foreground RM (97~~rad~m$^{-2}$) filament extending across the northern wing shell explains the different polarization angles in the northern wing shock shell 
in the Urumqi 4.8~GHz map of Fig.~\ref{6cmpc}. Based on the polarization angle deviation from the alignment with the shocked shell at 4.8~GHz,
we can roughly estimate the RM in the wing region.  
In the outer northern wing shock-shell of VRO~42.05.01, the polarization angle deviates less than 20$\degr$, indicating a lower RM of 80~rad~m$^{-2}$. 
After subtracting the foreground contribution, the RM in the northern shell is about 20~rad~m$^{-2}$.
The polarization angle deviation in the southern wing shock-shell is small.
The high polarization percentage there indicates enhanced regular magnetic field with less random magnetic field than other regions.

\section{Summary}
We have obtained the polarization data cube of the SNR VRO~42.05.01 at 1240~MHz using the FAST radio telescope and applied
RM-Synthesis to derive the Faraday depth spectrum. Only the ``shell'' in the low RM region shows weak polarization emission. 
The peak Faraday depth map shows a large foreground RM region of 60~rad~m$^{-2}$ extending along the wing section, which
coincides with a large-scale HI shell at $-20$~km~s$^{-1}$. The two depolarization patches in the wing region with RM of 97 and 55~rad~m$^{-2}$
coincide with two HI structures in the HI shell.
Faraday screen model fitting on the CGPS 1420~MHz full-scale polarization data reveals 
that it lies at 2.2-2.6~kpc in the Perseus arm with an enhanced regular magnetic field.
The highly piled-up magnetic field indicates that the HI shell at $-20$~km~s$^{-1}$ could originate from an evolved old SNR.

\begin{acknowledgements}
We thank the journal referee for critical reading and valuable comments to improve the paper.
We acknowledge the support from the National Key R\&D Program of China (2018YFE0202900), and the help from FAST colleagues for using the server.
The research presented in this paper has used data from the Canadian Galactic Plane Survey, a Canadian project with international partners, 
supported by the Natural Sciences and Engineering Research Council.

\end{acknowledgements}

\bibliographystyle{aa}

\bibliography{G166.bib}

\end{document}